\documentclass[reprint,secnumarabic, aps, prd]{revtex4-2}

\newcommand{\env}[1]{\texttt{#1}}
\usepackage{graphicx}
\usepackage{gensymb}
\usepackage{amssymb}
\usepackage{amsmath}

\begin{document}

\title{Orientation-sensed Optomechanical Accelerometers based on Exceptional Points}%

\author{Rodion Kononchuk}%
\email[]{rkononchuk@wesleyan.edu}
\affiliation{Wave Transport in Complex Systems Lab, Department of Physics, Wesleyan University, Middletown CT-06457, USA}

\author{Tsampikos Kottos}%
\affiliation{Wave Transport in Complex Systems Lab, Department of Physics, Wesleyan University, Middletown CT-06457, USA}

\date{January 2020}%
\begin{abstract}
	Abstract: We propose an optomechanical design, consisting of a parity-time symmetric multilayer structure tuned at exceptional-point degeneracy (EPD), with an adjustable layer that is coupled to micromechanical springs. The deflections of this layer in response to accelerations $\alpha$, lead to square-root resonance detuning.$\Delta\omega \equiv \omega -\omega_{EPD} = \sqrt{\alpha}$ - thus dramatically enhancing the probe of ultra-small accelerations $\alpha \ll 1$. Our design is scalable and can, in principle, support higher $N$-th order EPDs with sensitivity $\Delta\omega \propto \sqrt[n]{\alpha}$. It also provides a pathway towards a new generation of on-chip hypersensitive accelerometers and vibrometers.
\end{abstract}
\maketitle

\section{Introduction}
The monitoring of directional acceleration is essential for a variety of technological applications ranging from navigation devices, gravity gradiometry and earthquake monitoring, to intruder detection, airbag deployment sensors in automobiles and consumer electronics protection \cite{1,2,3,4}. Depending on the application at hand, the requirements of the performance metrics of the various type of accelerometers vary. Most of the existing schemes, nevertheless, utilize a linear response principle to measure a constant acceleration. These measurement protocols involve a mounted test mass whose displacement, due to the applied acceleration, is typically proportional to the acceleration and it is sensed using capacitive \cite{5}, piezo-electric\cite{6}, tunnel current \cite{7}, or optical methods \cite{8, 9, 10, 11, 12, 13}. A bottleneck for all these schemes is the accuracy of the displacement sensor which constitutes a central figure of merit of the accelerometers. Needless to point-out that optical sensing schemes have been proven up to now superior (as far as resolution is concern) to all others. In fact, the recent developments in optomechanics have spur the realization of a new generation of photonic motion detectors that surpass the sensitivity limitations of current standards, with aims to approach fundamental limits set by quantum mechanics \cite{14, 15}. These systems rely on the coupling between mechanical motion and light, which is enhanced when using optical resonators such as a Fabry-Perot cavity \cite{16} or nanocavities in the form of photonic crystals \cite{12, 17}.  Nevertheless, the requirement for accurate displacement measurements of the test-mass remains a daunting task. 

In this paper we propose a novel class of on-chip optomechanical accelerometers, that utilize concepts from non-Hermitian wave-mechanics, in order to achieve enhanced displacement measurements of the test-mass. Specifically, we exploit the rich mathematical structures underlying the presence of non-Hermitian spectral degeneracies, known as exceptional points (EP). These are branch point singularities in the parameter space of a system at which $N$ eigenvalues and their associate eigenvectors coalesce and become degenerate. This class of degeneracies has spawn significant interst and developments in optics \cite{36}, electronics \cite{37} and acoustics \cite{38}.   As opposed to Hermitian degeneracies, where the modes are expanded around their degenerate point in a Taylor series with respect to a small perturbation $\epsilon$, here the appropriate expansion is the Puiseux series. In particular, for a system supporting an $N$-th order EP, the frequency shift in response to a small perturbation $\omega_n = \omega_{EP}+\sum_{m=1}^{\infty} {c_m^{(n)}\epsilon^{m/N}}$. Given that $\epsilon^{1/N}\gg\epsilon $ for $|\epsilon|\ll1$ such non-Hermitian response opens up new horizons in probing small perturbations \cite{18, 19, 20}. In fact, this concept has been already implemented for the development of a new generation of hypersensitive ring laser gyroscopes \cite{21, 22}.

\begin{figure}
	\includegraphics[width =8.6cm]{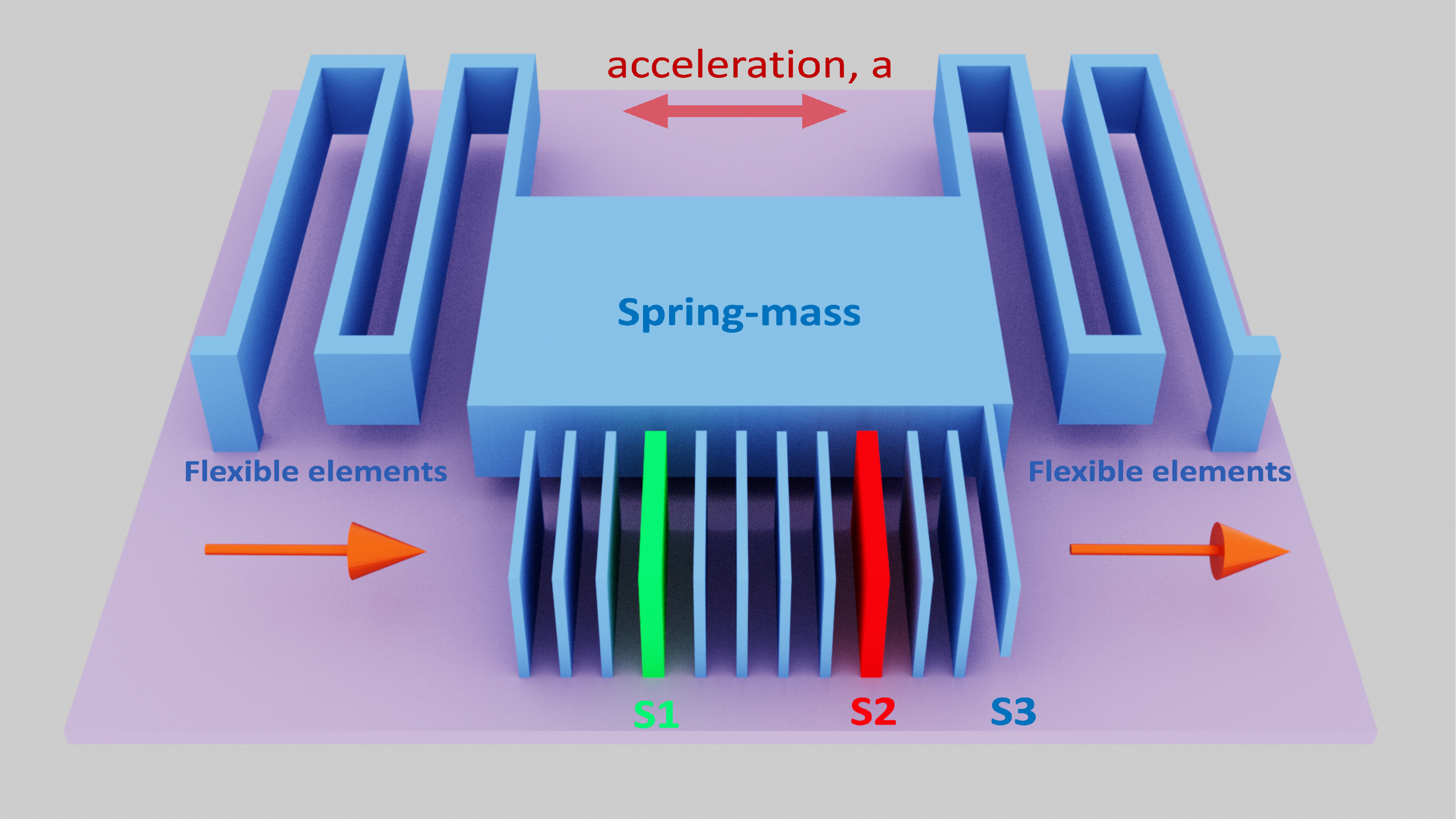}
	\caption{(a) Schematic of the proposed design of optomechanical accelerometer, consisting of pair of Fabry-Perot multilayer cavities with one of the cavities (S1) having losses $\gamma$  (green) and second (S2) having the equal amount of gain (red). The cavities are grown on silicon platform forming silicon/air multilayer. One of the silicon layers (S3) is attached to the spring-mass (blue) in order to sense the in-plane acceleration by perturbing the resonant frequency of one of the modes. The spring-mass system senses an in-plane directional acceleration. It consists of flexible silicon elements and a test mass connected to the silicon layer (S3)}
	\label{fig1}
\end{figure}

Here we propose an optomechanical EP-accelerometer consisting of a multilayer photonic crystal (PC) grown on a silicon platform  \cite{23, 24}, see Fig.~\ref{fig1}. The photonic crystal consists of quarter-wavelength silicon layers separated by air spacers. At symmetric positions with respect to the mirror plane of the structure we placed two optically identical silicon defect layers; each having half-wavelength width. One of these layers (S1) has loss, whose origin can be traced either to radiation losses or to absorption, while the other one (S2) has equal amount of gain. The gain can be introduced to the layer S2 by its partial thermal oxidation to SiO$_2$ \cite{23}, followed by injection of Erbium Er$^{3+}$ nanoparticles \cite{25} which are subsequently pumped optically \cite{26}. The total structure demonstrates a parity-time (PT) symmetry when the gain in layer S2 balances the amount of losses experienced by the S1 layer. In general, this PC supports two defect modes at distinct frequencies. When, however, the amount of gain/loss at each defect cavity matches the tunneling coupling between the (isolated) cavity modes, the system demonstrates an EP degeneracy. Our design assumes that in the absence of acceleration the system is at the EP.  We have completed our design, by further assuming that one of the silicon layers (S3) is attached to a spring mass; thus, acting as a test-mass that senses in-plane accelerations, see Fig.~\ref{fig1}. An in-plane acceleration in a direction parallel to the PC, will cause the adjustable silicon layer to shift closer (or further) from the rest of the stack, depending on the direction of the acceleration $\alpha$. Alternatively, one may also consider a design where the layer S3 is stationary (grown on the substrate) while the rest of the stack is attached to a test-mass and senses the acceleration. In both cases, the applied acceleration enforces a thickness variation of the air spacer layer between the silicon layer S3 and the rest of the stack. Consequently, this variation leads to a frequency detuning of the cavity formed by the defect layer S2, and to a subsequent lift of the EP degeneracy. In particular, one of the emerging modes becomes high-$Q$ and experiences a resonant shift $\Delta\omega\propto\sqrt{\alpha}$ 
while the other one remains at $\omega_{EP}$ and undergoes an abrupt overdamped transition. The proposed scheme is scalable, and the PC can be arranged to support an $N$-th order EP wih higher order sensitivity $\Delta\omega\propto\alpha^{1/N}$, by introducing $N$ PT-symmetric defects. Additionally, the sensitivity of the device can be increased by modifying the design in a way that few neighboring silicon layers are attached to the test mass and, thus move together further or closer to the rest of the stack once acceleration is applied. This will result in a thickness variation of the air spacer layer located closer to the defect silicon layer S2. Since the electric field is exponentially enchanced in the vicinity of the defect layers, in this case, the same thickness variation (applied acceleration) will lead to much stronger resonant shift detuning of the cavity formed by the layer S2, compared to the case discussed above.
\section{Design of Optomechanical  PC}

We consider a PC consisting of quarter-wave silicon (S) and air (A) layers with the arrangement (SA)$^3$S1(AS)$^4$AS2(AS)$^2$A1S3. The corresponding refractive indexes at an operational wavelength $\lambda\approx1550 nm$ are $n_S$=3.48 and $n_A$=1. The defect layers S1 and S2 are half-wave thick silicon layers with complex index of refraction $n_{S1}$=3.48+0.00217765i (losses) and $n_{S2}$=3.48- 0.00217765i (gain) respectively. S3 is an adjustable quarter-wavelength silicon layer which is attached to a test mass (Fig.~\ref{fig1}) and it is under-etched in order to avoid friction with the silicon surface - thus sensing the applied in-plane acceleration. Variations in its position due to an in-plane acceleration leads to changes in the thickness of the quarter-wavelength air spacer A1, and consequently to a resonant detuning of the S2 defect layer. In general  the thickness of the silicon layers S and S3 should satisfy $n_S\times d_S= \frac{(2\times N+1)\times\lambda_0}{4}, N\in\mathbb{Z}$, in order to support a Fabry-Perot mode at operatoinal wavelength $\lambda_0$, while the defect layers S1 and S2 should satisfy $n_{S1,2}\times d_{S1,2}= \frac{(2\times N)\times\lambda_0}{4}, N\in\mathbb{Z}$ \cite{23, 24}. This allows to make the layers as thick as it is required for a high-quality production process.

\begin{figure}
	\includegraphics[width =8.6cm]{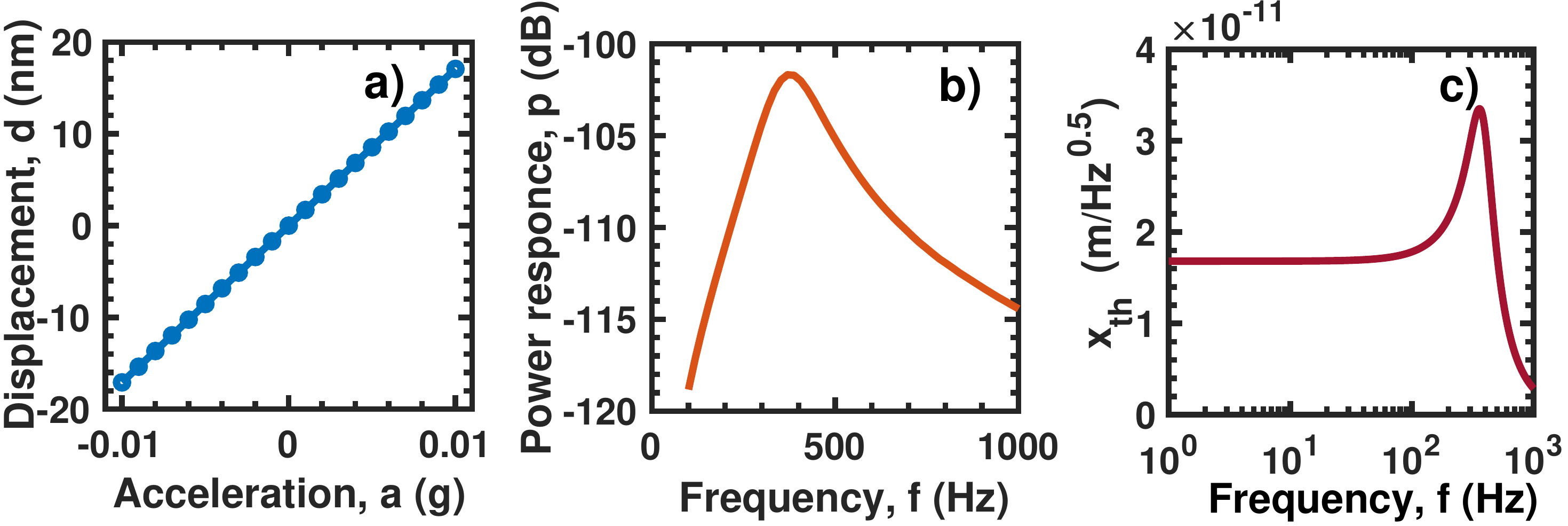}
	\caption{(a) Simulated displacement d of the layer S3 (together with the pair of two masses), as a function of applied acceleration $\alpha$. The inverse slope in this diagram provides the value of the (angular) natural frequency of the mechanical oscillator $\omega$=2.4 rad$\cdot$kHz. (b) The mechanical response of the oscillator of Fig.~\ref{fig1}. The mechanical linewidth is extracted from the resonant FWHM.  (c) The spectral density of the thermal noise displacement for the proposed accelerometer.	}
	\label{fig2}
\end{figure}

The micromechanical oscillator undergoing thermal and viscous damping due to the surrounding air has been designed using coupled structural mechanics and acoustic modules of COMSOL Multiphysics software \cite{27}. It consists of the test mass made of silicon (blue) connected with an adjustable silicon layer S3. The mass is coupled to a pair of flexible silicon elements (see Fig.~\ref{fig1}). The size of the silicon mass is 500 $\mu$m $\times$ 250 $\mu$m $\times$ 5 $\mu$m, while the flexible elements consist of four silicon springs connected in series (as shown in Fig.~\ref{fig1}) with dimensions 1 $\mu$m $\times$ 400 $\mu$m  $\times$ 5 $\mu$m, separated by 15 $\mu$m air gaps, responsible for thermal and viscous damping.
The main requirements of the design was to guarantee (i) a uniform displacement of the S3 layer for the range of accelerations that we have used, (ii) the validity of a linear response between the test-mass displacement and the acceleration in a low frequency  limit (see appendix)  i.e. $ \alpha = \omega_n^2 x_{res}$ (where $\omega_n$  is the angular natural frequency of the mechanical oscillator) and (iii) a low level of  the thermal noise. The displacement of the mass elements and of the silicon layer S3, as a function of an applied horizontal in-plane acceleration, is shown in Fig.~\ref{fig2}a and nicely demonstrates a linear response behavior which allows  us to extract 
mechanical resonant frequency $\omega_n$ i.e. $\omega_n \equiv \sqrt{\frac{k_{eff}}{m_{eff}}}$ where $m_{eff}\approx $ 2.8$\mu$g, is the effective mass and $ k_{eff}\approx$ 12.3 mN$\cdot$m$^{-1}$ is the effective spring constant. It is important to point out that small resonance frequencies $\omega_n$ are frequently incompatible with the necessity for high-speed operations. At the same time, large resonance frequencies $\omega_n$ lead to vanishingly small susceptibilities and therefore displacements. It is therefore imperative to develop measurements protocols that sense vanishingly small displacements.

Another important characteristic of the micromechanical oscillator is its mechanical $Q$-factor.  The latter dictates the response time $\tau_R=Q/2\omega_n$ of the oscillator to an external force. In Fig.~\ref{fig2}b we present a frequency domain simulation for the oscillator showing a power response $p=20\cdot\log_{10}(|\omega d|)$  , which confirms the value of  $\omega_n$ and allows us to extract the value of $Q\equiv \omega_n/\Delta \omega_{FWHM}$=1.99, where $\Delta \omega_{FWHM}$ is the full width at half maximum (FWHM)  and therefore $\tau_R$ = 3.96$\cdot$10$^{-4}$  sec. The knowledge of these parameters provides an estimate of the lower bound sensitivity and noise effects on acceleration measurements. In particular, the fundamental resolution limit in acceleration measurements is typically set by the presence of a thermal Brownian motion of the test-mass. The latter is quantified by the thermal noise equivalent acceleration $\alpha_{th}$ which provides the lower bound in acceleration measurements \cite{28, 29}. It is given by (see appendix)

\begin{equation}
\alpha_{th} = \sqrt{\frac{4k_bT\omega_n}{m\cdot Q}} ~~~\left[\frac{g}{\sqrt{Hz}}\right];
\label{e1}
\end{equation}
where $k_b$  is the Boltzmann constant and $T$ = 293.15 K is the temperature. In fact, there are two other noise sources related with (i) a readout noise ($\alpha_{det}$) and (ii) other noise sources ($\alpha_{add}$)  such as the ones associated with the use of the interrogating laser and the electronics used for the measurement process. However, these two types of noise sources can be minimized, leaving $\alpha_{th}$ the main source that is responsible for bounding the resolution in acceleration measurements \cite{12,13}. In fact, from Eq.~(\ref{e1}) we see that a high resolution at room temperatures, requires high values for the product $m\cdot Q$ i.e. both large mass and high mechanical $Q$-factor. Using the extracted value of $Q$ we find from Eq.~(\ref{e1}) that $\alpha_{th}\approx$10 $\mu$g/$\sqrt{Hz}$. We conclude this analysis by pointing out that in our calculations for the $Q$ -factor we took into consideration only the thermal and viscous damping due to the surrounding air. Contributions like the non-contact friction (in vacuum) associated with interactions between the adjustable layer S3 and the silicon surface (or the silicon layer adjacent to S3) are typically smaller than the air damping, and therefore have been disregarded from our analysis. 
Using the above analysis, we can further evaluate the spectral density of the thermal noise displacement of the proposed accelerometer. It is given by (see appendix)
\begin{equation}
x_{th}(\omega) = \sqrt{\frac{4k_bT\omega_n}{m\cdot Q \left( {\left(\omega_n^2-\omega^2\right)}^2 +\left(\frac{\omega_n \omega}{Q}\right)^2 \right)}};
\label{e2}
\end{equation}
and is plotted on Fig. \ref{fig2}c.  It shows almost flat spectral noise density at frequencies $f \leq 100 $ Hz, implying that maximum bandwidth of the proposed accelerometer is about 100 Hz. The associated RMS noise acceleration, measured within this bandwidth $\Delta f =100 $ Hz (low pass filter) will be $\langle \alpha_{noise} \rangle= \alpha_{th}\cdot \sqrt{\Delta f}=100$ $\mu g$ (see appendix), which sets the noise floor level (see inset on Fig. 4c).

\section{PCs with Exceptional Point Degeneracy}

First we will investigate the existence of EP degeneracy for the PT-symmetric PC. To this end, we analyze its transmission spectrum $T(\omega)$, in the absence of acceleration, using transfer matrices. The latter connects the amplitudes of forward and backward propagating waves between two consequent layers (and between the first/last layer and the left/right ambient space). In particular, a time-harmonic field of frequency $\omega$, satisfies the following Helmholtz equation inside each spatial domain $0\leq  x_j\leq x_j+L_j$  ($L_j$  is the width of the $j$-th layer)

\begin{equation}
\frac{d^2 E(x)}{dx^2}+\left( \frac{\omega}{c}\right)^2 \varepsilon(x)E(x) = 0;
\label{e3}
\end{equation}
At the$j$-th layer, Eq.~(\ref{e3}) admits solutions of the form
 $E^{(j)} (x)=E_f^{(j)} e^{in_j kx}+E_b^{(j)} e^{-in_j kx}$, where $k=\omega/c$ is the free space (air) wavevector. Similarly, on the left and right of the PC, Eq.~(\ref{e3}) admits the solution $E^{(L/R)} (x)=E_f^{(L/R)} e^{ikx}+E_b^{(L/R)} e^{-ikx}$. The continuity of the field and its derivative at the interface between two layers (or a layer and the left-right ambient space) is expressed in terms of the total transfer matrix $M$ which connects the forward and backward amplitudes on the left ($L$) and right ($R$) of the PC:

\begin{equation}
\left( \begin{array}{ccc}
E_f^{(R)}   \\
E_b^{(R)}  \end{array} \right)  = 
\left( \begin{array}{ccc}
M_{11} & M_{12}  \\
M_{21} & M_{22}  \end{array} \right) 
\left( \begin{array}{ccc}
E_f^{(L)}   \\
E_b^{(L)}  \end{array} \right),
M = \prod_{j=0}^{q}M_{j};
\label{e4}
\end{equation}
where $q$ is the total number of layers. The single-layer transfer matrix $M_j$ connects the field amplitudes of the $j$-th and the $(j+1)$-th layers i.e. $\left(E_f^{(j+1)},E_b^{(j+1)}\right)^T=M_j \left(E_f^{(j)},E_b^{(j)} \right)^T$. Thus, the transfer matrix approach allows us also to construct the field $E^{(j)}(z)$ at each layer, provided that appropriate scattering boundary conditions are imposed. The latter, for a left incident wave, take the form $\left(E_f^{(R)},E_b^{(R)} \right)^T=\left(1,0\right)^T$. It is easy to show that $T=\left|\frac{1}{M_{22}} \right|^2$, and $R=\left|\frac{M_{21}}{M_{22}} \right|^2$ \cite{30, 31}.

\begin{figure}
\includegraphics[width =8.6cm]{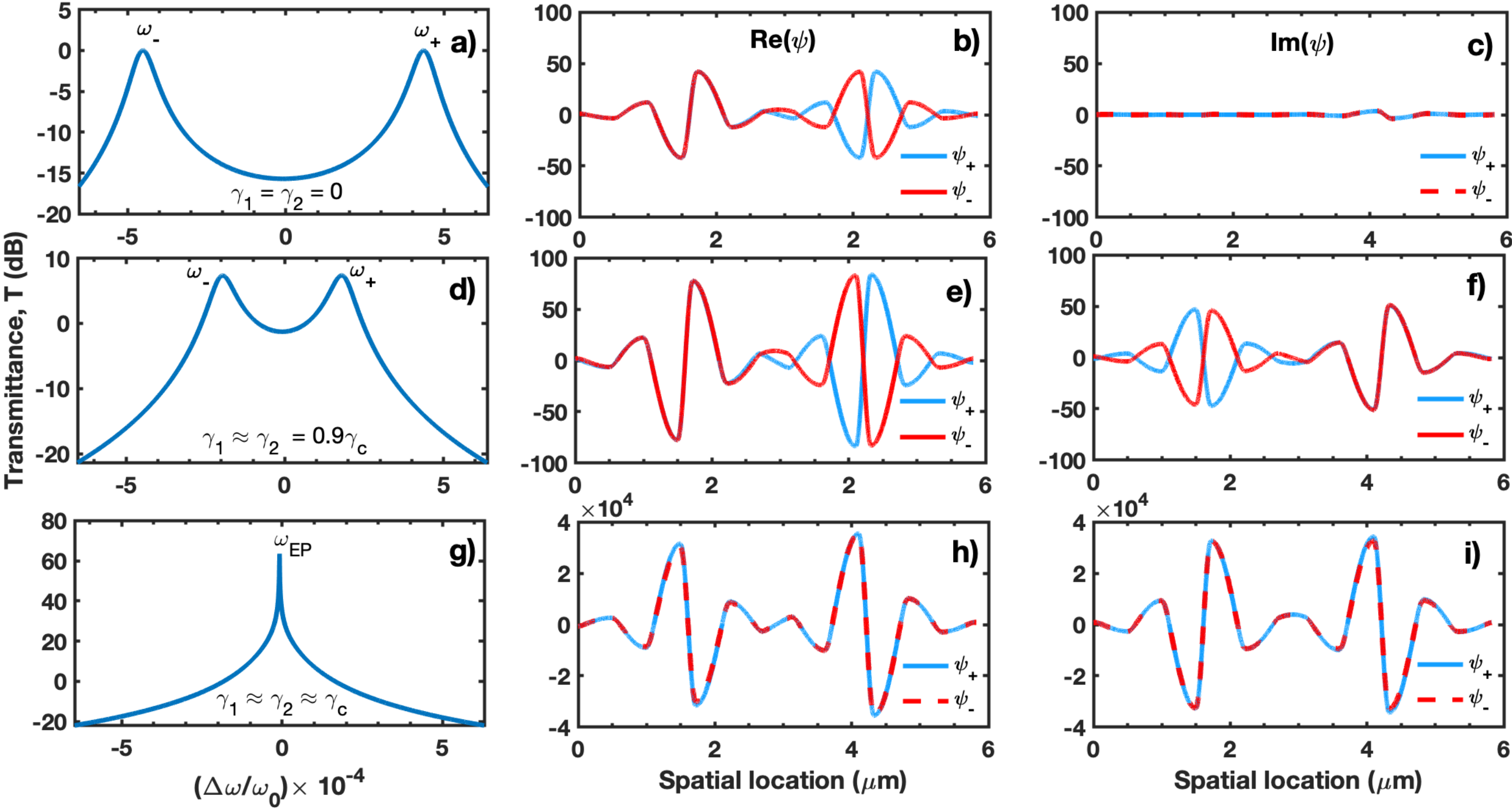}
\caption{ (a) Simulated transmittance spectra of the multilayer, without gain and loss ($\gamma_1 = \gamma_2 = 0$) in the vicinity of the resonant frequency $\omega_0$.   Corresponded spatial profile within the multilayer of the real (b) and imaginary (c) parts of forward propagated wave functions $\psi_+$ and $\psi_-$. (d) Transmittance spectra of the multilayer, close to the EP condition ($\gamma_1 \approx \gamma_2  = 0.9 \gamma_c$). Corresponded spatial profile within the multilayer of the real (e) and imaginary (f) parts of forward propagated wave functions $\psi_+$ and $\psi_-$. (g) Transmittance spectra of the multilayer, very close to the EP condition ($\gamma_1 \approx \gamma_2 \approx \gamma_c$). Corresponded spatial profile within the multilayer of the real (h) and imaginary (i) parts of forward propagated wave functions$\psi_+$ and $\psi_-$.
}
\label{fig3}
\end{figure}
We first analyze the transmission spectrum $T(\omega)$  in the case where the two defect cavities S1, and S2 are identical i.e. $Im(n_{S1} )=Im(n_{S2} )=0$. In this case the cavities support pairs of identical cavity modes which are degenerate at $\omega_0$. When the two cavities are in the proximity of one-another, the cavity modes supported by each individual cavity, are coupled together forming two “supermodes” with eigenfrequencies $\omega_+$  and $\omega_-$ (see Figs.~\ref{fig3}a,b,c).  In particular, the frequency splitting is controlled by the number of layers between the two defect cavities. In Fig.~\ref{fig3}a we show the transmission spectrum of the PC with the two identical defect layers. It shows two resonant peaks (associated with each of the two defect supermodes) forming close to the center of the band-gap. The associated supermodes $\psi_+$ and $\psi_-$ are a linear (symmetric and antisymmetric) superposition of the individual cavity modes  (see Figs.~\ref{fig3}b,c). When we introduce gain/loss to the two defect layers i.e. $ Im(n_{S1} )=-Im(n_{S2} )\neq0$ the two resonant peaks of $T(\omega))$ start approaching one-another (see Fig.~\ref{fig3}d). This is a consequence of the PT-symmetry which essentially renormalizes the coupling between the cavity modes, thus promoting the formation of an EP degeneracy. At the same time the supermodes remain also eigenmodes of the PT-symmetric operator (albeit they are now complexed valued), see Figs.~\ref{fig3}e,f. When the gain/loss strength becomes equal to the coupling strength between the individual cavities, the two resonant peaks coalesce and form an EP degenerate pair at $\omega = \omega_{EP} =\omega_0$, see Fig.~\ref{fig3}g. At the same time the supermodes of the system become degenerate (Figs.~\ref{fig3}h,i).

\section{EP-based optomechanical accelerometers }

\begin{figure}
	\includegraphics[width =8.6cm]{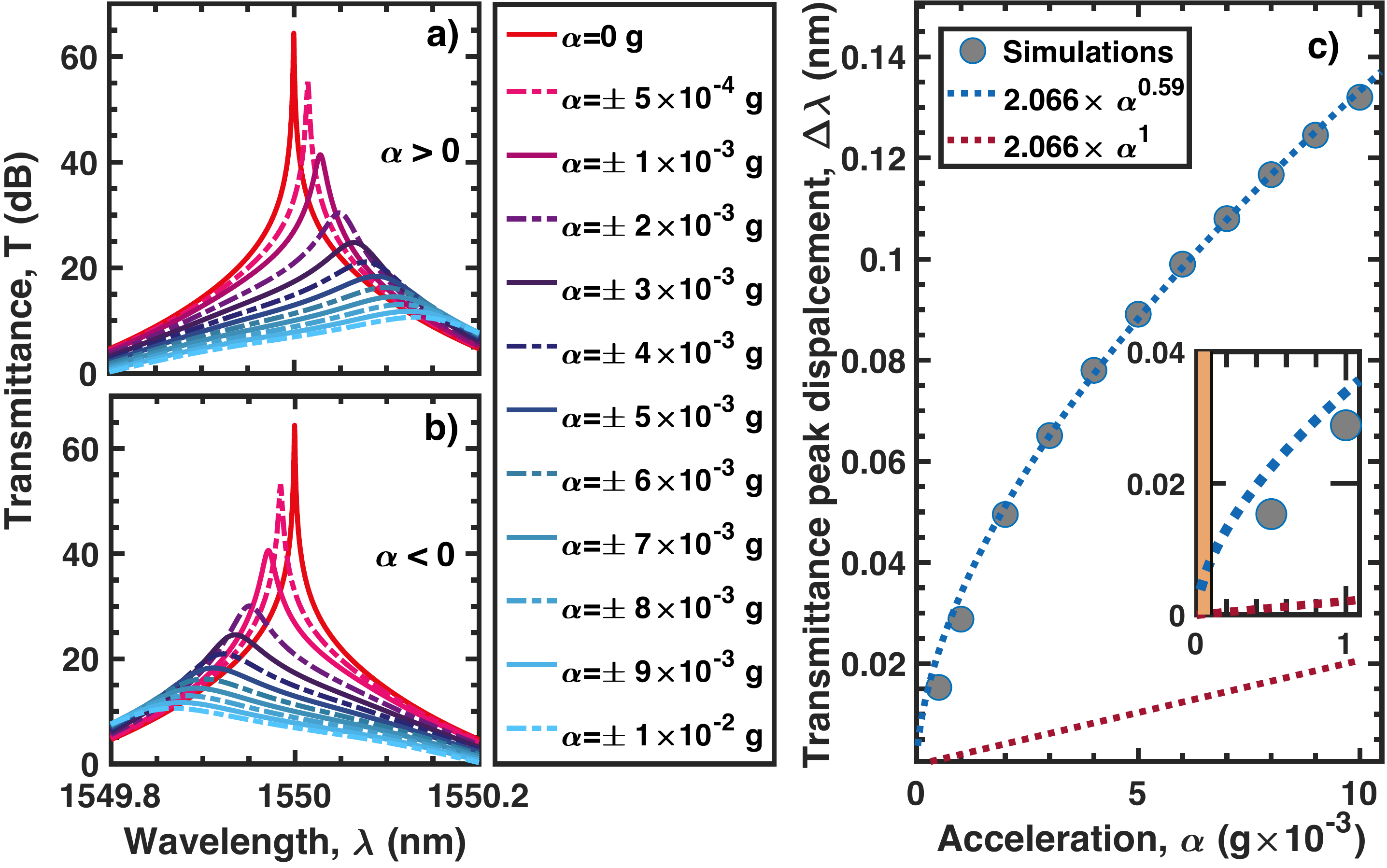}
	\caption{(a) Simulated transmittance spectra of the optomechanical accelerometer schematically drown on Fig.~\ref{fig1} with integrated spring-mass system for various small values of applied right-directed acceleration ($\alpha>0$). (b) Simulated transmittance spectra for various small values of left directed acceleration ($\alpha<0$). (c) Transmittance peak displacement $\Delta \lambda$ from the E.P. spectral degeneracy obtained from panel (a) as a function of applied acceleration $\alpha$ (black square dots) and its best fit with the function $2.066\cdot \alpha^{0.59}$  (dashed blue line). Linear function $2.066\cdot \alpha$  (dashed red line) is added for comparison. The orange region on the inset indicates the RMS noise acceleration, (i.e. floor noise level).
	}
	\label{fig4}
\end{figure}

Next, we investigate the consequences of the acceleration $\alpha$ at the EP degeneracy. We note that for non-zero acceleration $\alpha \neq0$,  Eq.~(\ref{e3}) has to be solved taking into account the adjusted equilibrium position of the S3 layer (see Fig.~\ref{fig2}a) and, consequently, the modified width $L_{A1}$ of the last air spacer (A1). The latter can increase or decrease (with respect to its value for $\alpha=0$), depending on the orientation of the acceleration. Figs.~\ref{fig4}a,b show the transmittance $T(\lambda)$  for various values of the applied acceleration $\alpha$. We observe a (red/blue) shift of the resonant peak of $T(\lambda)$  depending on the direction of the acceleration. In Fig.~\ref{fig4}c we present a panorama of these resonant shifts versus the acceleration. A fractional power-law behavior of the peak wavelength $\lambda_{max}$ versus $\alpha$ is evident i.e. $\lambda_{max} \approx 2 \sqrt{\alpha} $ [g] . For comparison purposes, we also plot at the same figure a standard linear dependence $\lambda_{max} \approx2  \alpha$ [g] occurring in case of typical optomechanical accelerometers based on linear response. This comparison implies that the proposed EP optomechanical accelerometers has a dramatic sensitivity improvement as far as the displacement of the test-mass is concern. The inset on Fig.~\ref{fig4}c illustrates the floor noise level (orange region) of the measured accelerations calculated from Eq.~(\ref{e1}).

The theoretical analysis of the square-root response can be performed using a CMT. Since the cavity defect modes are located in the band-gap and are well separated from the rest of the Fabry-Perot modes, we can eliminate them from our consideration. The two interacting (via tunneling) defect cavity modes are described by the following 2$\times$2 effective Hamiltonian 

\begin{equation}
H_0 = \ \left( \begin{array}{ccc}
\omega_0-i\gamma_1 & \kappa  \\
\kappa &\omega_0+i\gamma_2  \end{array} \right);
\label{e5}
\end{equation} 
where $\omega_0$ is a eigenfrequency of the individual stand-alone cavity mode, $\gamma$  is their linewidth due to the presence of the loss/gain mechanisms at the first and second cavities respectively and $\kappa$ is the coupling between the two cavities due to tunneling across the in-between layers of the PC. One can diagonalize Eq.~(\ref{e5}) and evaluate the eigenfrequencies of the coupled cavity system:

\begin{equation}
\omega_{\pm} = \omega_0 \pm \sqrt{\kappa^2-\gamma^2}
\label{e6}
\end{equation} 
with corresponding eigenvectors $\upsilon_{\pm}=\left(\frac{-i\gamma \pm  \sqrt{\kappa^2-\gamma^2}}{\kappa},1\right)^T$. The above equations imply that the two modes become degenerate $\omega_{\pm}=\omega_{EP}=\omega_0$ (EP degeneracy), when $\kappa_{EP}\equiv\kappa=\gamma$. We consider that the two cavity modes of the PC are satisfying the EP condition.

Next we assume that the resonant frequency of the second Fabry-Perot cavity (the one that it is closer to the edge layer S3) is detuned from its original eigenfrequency value $\omega_0$ by   $\epsilon$. The physical origin of the detuning is associated with the displacement of the adjustable layer S3 which “deforms” the right mirror; thus, affecting the right-cavity modes: A right-directional acceleration will lead to a left-directed fictitious force and a subsequent displacement of the S3 layer towards the PC. This positional variation decreases the width of the A1 air spacer and produces a blue shift in the cavity frequencies $\omega_0\to\omega_0+\epsilon$. An opposite displacement will produce a red-shift to the cavity frequencies $\omega_0\to\omega_0-\epsilon$ (we assume $\epsilon>0$). This process is modeled by the following modified effective Hamiltonian: 

\begin{equation}
H_{\epsilon} = \ \left( \begin{array}{ccc}
\omega_0-i\gamma_1 & \gamma  \\
\gamma &\omega_0+\epsilon+i\gamma_2  \end{array} \right);
\label{e7}
\end{equation} 
where we have assumed that the EP condition $\kappa = \gamma$ is satisfied. The new set of perturbed eigenfrequencies are:

\begin{equation}
\omega_{\pm} = \omega_{EP}+\frac{\epsilon}{2} \pm \frac{\sqrt{\epsilon^2+4i\gamma\epsilon}}{2};
\label{e8}
\end{equation} 
For small frequency detuning $\epsilon \ll \gamma$ we have 
\begin{equation}
\left\{
\begin{array}{cc}
\omega_+ \approx \omega_{EP}+\frac{\sqrt{2}}{2}(1+i)\sqrt{\epsilon}\\
\omega_- \approx \omega_{EP}-\frac{\sqrt{2}}{2}(1+i)\sqrt{\epsilon}
\end{array}\right.
\label{e9}
\end{equation}
while for large detuning  $\epsilon\gg\gamma$ we recover a standard linear shift i.e. 

\begin{equation}
\left\{
\begin{array}{cc}
\omega_+ \approx \omega_{EP}+\epsilon+i\gamma\\
\omega_- \approx \omega_{EP}-i\gamma
\end{array}\right.
\label{e10}
\end{equation}

\begin{figure}
	\includegraphics[width =7cm]{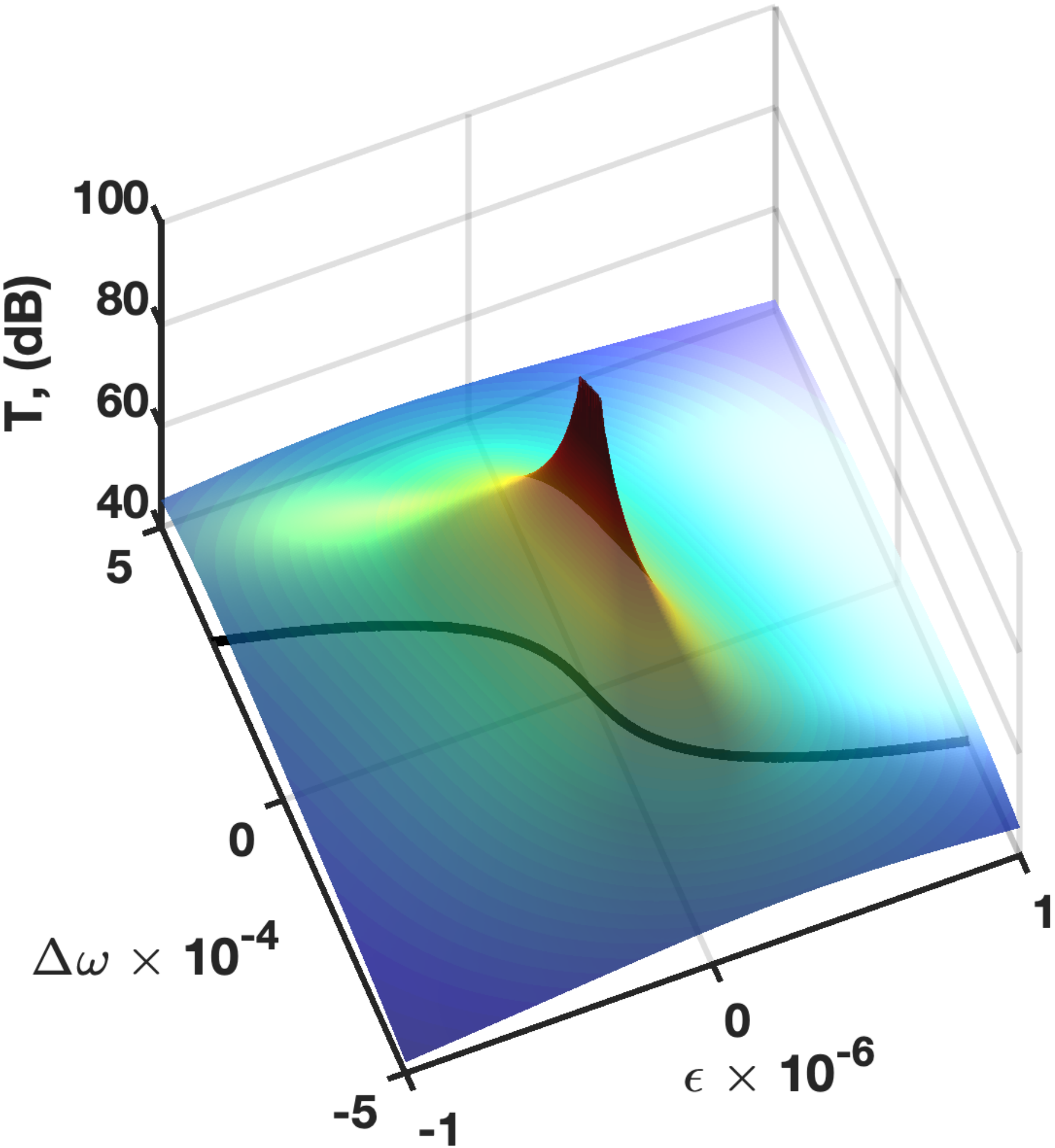}
	\caption{Theoretical value of transmittance $T$ (in dBs) versus frequency detuning $\Delta \omega=\omega_{EP}-\omega$ from the EP and resonant perturbation $\epsilon$. In calculations the $\kappa=\gamma=0.3$, $w_e=10^{-4}$. The solid line indicates the resonance detuning (at peak transmittance) for each $\epsilon$-value. A square-root behavior $\Delta \omega (T_{peak}) \propto \sqrt{\epsilon}$ is evident.
	}
	\label{fig5}
\end{figure}
Our interest, of course, will focus in the parametric domain where the conditions (i.e. $\epsilon \ll \gamma$) for the validity of Eq.~(\ref{e9}) are satisfied. In fact, this enhanced response of the eigenfrequency shifts carry over to the scattering domain where it surface as a shift of the poles $\omega^S$ of the scattering matrix $S$. The associated 2$\times$2 scattering matrix takes the form \cite{32}
\begin{equation}
S= -I-iW^T\frac{1}{H_\epsilon-\omega I-\frac{i}{2}WW^T}W;~~~W_{nm}=\sqrt{2w_e}\delta_{nm}
\label{e11}
\end{equation}
where $I$ is the 2$\times$2 identity matrix and $W$ is the coupling matrix that connects the two modes to the free space with a coupling strength $w_e$. The poles of the scattering matrix are found easily from the secular equation $det\left[H_\epsilon-\frac{i}{2} WW^T-\omega I\right]=0$. We get 
\begin{equation}
\omega_{\pm}^S = \omega_{EP}+\frac{\epsilon-2iw_e}{2} \pm \frac{\sqrt{\epsilon^2+4i\gamma\epsilon}}{2};
\label{e12}
\end{equation}
which in the limit of $w_e\to 0$ collapses to the eigenfrequencies of the dimer systems Eq.~(\ref{e8}). In fact, both $\omega_{\pm}$ in Eq.~(\ref{e8}) and the poles $\omega_{\pm}^S$ in Eq.~(\ref{e12}) demonstrate the same qualitative features on their dependence on $\epsilon$. Let us analyze these features by referring, for simplicity, to the poles – the latter being directly relevant to the proposed acceleration measurement scheme. For small $\epsilon \ll \gamma$ both poles are shifted away from $\omega_{EP}$  in opposite direction with a rate which is proportional to $\sqrt{\gamma \epsilon}$. At the same time  $Im(\omega_+^S )$ becomes positive and larger, leading to a narrow, high-$Q$, transmittance peak. In contrast, the $Im(\omega_-^S )$ broadens with the same rate $\sqrt{\gamma \epsilon}$ leading to a low-$Q$ resonant transmission peak (see Fig.~\ref{fig5}).  In fact, the linewidth broadening of the $\omega_-^S$ - pole occurs at the same rate as a shift of the resonant frequencies $\omega_+^S$. As a consequence, these resonant modes have indistinguishable peaks in the transmission spectrum $T(\omega)$. Once the detuning becomes large enough $\epsilon \gg \gamma$ the $\omega_-^S$  pole moves back to $\omega_{EP}$ and its position remains unchanged with further increase of $\epsilon$. Its linewidth $Im(\omega_-^S )$  continues to broaden (due to negative imaginary component $-i\gamma$) leading to a suppression of the transmittance (see Fig.~\ref{e5}). In contrast, the other pole $\omega_+^S$ is shifted further away from the $\omega_{EP}$  with a rate that is proportional to $\epsilon$ and at the same time it becomes narrower (due to positive imaginary component $+i\gamma$).  These features are nicely reflected in the transmission spectrum of our structure, see Fig.~\ref{fig5}. The latter is analytically evaluated from Eq.~(\ref{e11}): 
\begin{equation}
\begin{split}
T(\omega) \equiv \left|S_{12}(\omega)\right|^2= \\
=\left|  \frac{2i\kappa w_e}{\left( \Delta\omega-i(\gamma+w_e) \right)\left(\Delta\omega +\epsilon+i(\gamma-w_e)\right)-\kappa^2} \right|^2
\label{e13}
\end{split}
\end{equation}
where $\Delta\omega=\omega_{EP}-\omega$  while at EP. $\kappa = \gamma$ and resonant frequencies are given by Eq.~(\ref{e12})
.
Finally, we point out that  the rapid suppression of one of the transmission peaks (associated with $\omega_-^S$) and the simultaneous formation of a high-$Q$ resonant mode at $\omega_+^S$ results in a clearly resolved transmittance peaks in the spectrum without the concern of resolution digression due to (partially) overlapping resonances. As opposed to recent proposals that utilize EP for the realization of ring-laser gyroscopes, our design is operating below the lasing threshold and therefore must not be affected by quantum-noise effects \cite{33, 34, 35}. In contrast to the ring-laser resonators, the Fabry-Perot multilayer supports only one high-$Q$ defect mode within the band gap (the other mode is overdamped and suppressed). This feature leads to an improved spectral resolution of the devise and thus to the precision of the measurement. Another important advantage of our design, is its tolerance to global temperature variations of the multilayer: In particular, a temperature variation that induces a change in the thickness or the refractive index of the silicon layers, will affect uniformly the whole multilayer, producing the same shift in both resonant frequencies $\Delta\omega_+ =\Delta\omega_- $. Consequently, such thermally-induced variations will not shift the system from the EP -  as opposed to perturbations which affect only one of the modes: $\Delta\omega_+ =0$, $\Delta\omega_-\neq 0$. Such behavior provides additional degree of robustness of the device against parasitic (thermal) perturbations.  
\section{Conclusion}
In conclusion, we proposed a novel conceptual design of an optomechanical accelerometer based on two coupled Fabry-Perot cavities with a balanced amount of gain and loss. The structure is tuned to be at an EP in the absence of any acceleration. One of the layers is supposed to be attached to spring mass in order to sense the in-plane acceleration. The applied acceleration results in a displacement of the silicon layer with respect to the rest of the stack. This displacement is proportional to the acceleration and triggers a resonant frequency detuning of the high-$Q$ mode, and thus removal of the EP degeneracy. Using a theoretical model based on coupled mode theory, we have described the effects of the acceleration in the two defect modes that form the EP. We found that the resonant peak of the transmittance shifts in a square-root manner as a response to the displacement (and thus acceleration) of the test mass. Detail simulations confirmed this enhanced sensitivity. Our proposed scheme is superior to conventional acceleration sensors for which the resonant transmission peak shifts proportionally to the test-mass displacement (and thus to the acceleration). 

\begin{acknowledgments}
This research was supported by Office of Naval Research via grant
 N00014-19-1-2480 and  by Air Force Office of Scientific Research via grant  FA9550-10-1-0433
\end{acknowledgments}
\appendix

\section {Responce function }
 Consider a damped mechanical harmonic oscillator with mass $m$, spring constant $k$, natural frequency $\omega_n\equiv\sqrt{\frac{k}{m}}$,  attenuation parameter $\gamma$ and applied external force $F_{appl}$  . Such a system is described by the following differential equation of motion 
 \begin{equation}
 m\ddot{x}+m\gamma \dot x+m\omega_n^2 x = F_{appl};
 \label{a1}
 \end{equation}
 which after Fourier transform is written as 
 
 \begin{equation}
- \omega x(\omega)+i\gamma x(\omega)+\omega_n^2x(\omega)= \frac{F_{appl}(\omega)}{m}.
 \label{a2}
 \end{equation}
 Substituting $\alpha(\omega) = F_{appl}(\omega)/m$, one finally has
 
 \begin{equation}
 x(\omega) = \chi(\omega)\alpha(\omega);
 \label{a3}
\end{equation}
 where
 \begin{equation}
\chi(\omega) = \frac{1}{\omega_n^2-\omega^2+i\frac{\omega_n \omega}{Q}}
 \label{a4}
 \end{equation}
is the susceptibility of the mechanical oscillator which defines the response of the oscillator to external frequency dependent excitations associated with a force $F_{appl}(\omega)$. Above, $Q=  \omega_n/\gamma $ is the quality factor of the oscillator. From Eq.~(\ref{a4}) it is seen that in the low frequency limit ($\omega \ll \omega_n$) the susceptibility is $ \chi(\omega \ll \omega_n) \propto \frac{1}{\omega_n^2}$, corresponding to a flat frequency response of the mechanical oscillator. At the resonance frequency $\omega_n$  the response of the oscillator is enchanced by the $Q$ factor:  $\chi(\omega_n)=-i\frac{Q}{\omega_n^2}$. At the same time, at frequencies much higher than the natural frequency, the susceptibility decays in a square manner as a function of frequency: $\chi(\omega\gg \omega_n) \propto\frac{1}{\omega^2}$, leading to a very weak response of the mechanical system to the excitations with frequencies higher than the resonant one.  Such functional behavior leads to the conclusion that an accelerometer, in order to be able to measure the accelerations with different frequencies with the same efficiency (i.e. to be a broad-band sensor) has to operate well below the natural frequency $\omega_n$ where susceptibility  $\chi$ provides a flat frequency response.   In this case, on the other hand, the natural frequency defines the sensitivity of the accelerometer, leading to a fundamental trade-off between the sensitivity of the mechanical accelerometer sensitivity and its bandwidth. 

\section {Thermal noise}
Any mechanical oscillator is subject to parasitic random oscillations due to a Brownian motion within the surrounding media. This random excitation due to thermal noise is translated into a force $F_{noise}=m\alpha_{noise}$, which sets the fundamental resolution limit of an accelerometer. The mechanical system which is subject to random excitations due to Brownian motion will follow the differential equation
\begin{equation}
im\nu(\omega)+m\gamma\nu(\omega)-i\frac{k}{\omega}\nu(\omega)=F_{noise}(\omega);
\label{b1}
\end{equation}
which can be obtained from Eq.~(\ref{a1}) by substituting $\nu\equiv \dot{x}$ and $F_{noise}=F_{appl}$. Substituting the expressions $Q=  \omega_0/\gamma $ and $ f=\omega/(2\pi)$ in the above equation leads to the following expression for the velocity 
\begin{equation}
\nu(f) =\frac{F_{noise}(f)}{\gamma m\left( 1+iQ\left(\frac{f}{f_n}-\frac{f_n}{f} \right)\right)}.
\label{b2}
\end{equation}
The average (over spectrum) force due to thermal oscillations is described by the root mean square (RMS) value $\langle F_{noise}\rangle$ where the overbar indicates a spectral averaging. We have 
\begin{equation}
\langle F_{noise}\rangle^2 = \int_{f_1}^{f_1+\Delta f} \frac{F_{noise}^2(f)}{\Delta f}df= \int_{f_1}^{f_1+\Delta f} \overline {F_{noise}^2(f)}df.
\label{b3}
\end{equation}
Assuming that the thermal noise is white noise we can substitute the force spectral density with a constant i.e. $\overline {F_{noise}^2(f)}=Const$. Therefore, Eq.~(\ref{b3}) can be further simplified to 
\begin{equation}
\langle F_{noise}\rangle^2=\overline {F_{noise}} \cdot \Delta f
\label{b4}
\end{equation}
where  $\Delta f$  is the bandwidth of the measured accelerations i.e. the inverse of time averaging of the processed signal, meaning that noise signal with frequencies above $f_1+\Delta f$  and below $f_1$ will be averaged to zero, and thus will not affect the measurements. Then, from Eq.~(\ref{b2}) one obtains
\begin{equation}
\begin{split}
\overline{\nu^2} =\frac{\overline{F_{noise}^2}}{\left|\gamma m\left( 1+iQ\left(\frac{f}{f_n}-\frac{f_n}{f} \right)\right)\right|^2} =\\
=\frac{\overline{F_{noise}^2}}{\gamma^2 m^2\left( 1+Q^2\left(\frac{f}{f_n}-\frac{f_n}{f} \right)^2\right)} .
\label{b5}
\end{split}
\end{equation}
The mean kinetic energy stored in the mechanical system is defined as $E=\frac{1}{2} m \langle \nu \rangle^2$, where $\langle \nu \rangle $  is the RMS velocity of the oscillator due to thermal noise.  On the other hand, the total kinetic energy of the thermal noise oscillations from the all frequencies ($f\in[0,\infty]$) is $E=\frac{1}{2} k_b T$, where $k_b$ is the Boltzmann constant and $T$  is the temperature. Equipartition theorem leads to 
\begin{equation}
E= \frac{1}{2} k_b T =\frac{1}{2} m \langle \nu \rangle^2 = \frac{1}{2} m \int_{0}^{\infty} \overline{\nu^2} df.
\label{b6}
\end{equation}
Substituting Eq.~(\ref{b5}) into Eq.~(\ref{b6}) one can write
\begin{equation}
\frac{1}{2} k_b T = \frac{\overline{F_{noise}^2}}{4 \pi \gamma m } \int_{0}^{\infty} \frac{Q\cdot d\left(\frac{f}{f_n}\right)}{\left(1+Q^2\left(\frac{f}{f_n}-\frac{f_n}{f}\right)^2\right)}.
\label{b7}
\end{equation}
We can make further progress with the integration of Eq.~(\ref{b7}) by substituting $\frac{f}{f_n }= e^x
$. We get 
\begin{equation}
\frac{1}{2}k_b T = \frac{\overline{F_{noise}^2}}{8\gamma m};
\label{b8}
\end{equation}
leading to
\begin{equation}
\overline{F_{noise}^2} = 4 k_b T\gamma m.
\label{b9}
\end{equation}
Thus, the spectral density of the thermal noise equivalent acceleration $\alpha_{th} =\sqrt{\frac{\overline{F_{noise}^2}}{m^2}} $ can be written as
\begin{equation}
\alpha_{th} =\sqrt{\frac{4 k_b T \gamma}{m}} =  \sqrt{\frac{4 k_b T \omega_n}{m\cdot Q}}. 
\label{b10}
\end{equation}
Then from Eq.~(\ref{b4}) and Eq.~(\ref{b5}), the RMS acceleration due to the thermal noise is 
\begin{equation}
\langle \alpha_{noise} \rangle = \sqrt{\frac{\langle F_{noise}\rangle^2}{m^2}} = \sqrt{\frac{ \overline{ F_{noise}^2} \Delta f}{m^2}} = \alpha_{th}\sqrt{\Delta f},
\label{b11}
\end{equation}
which is the lower bound of the measured accelerations within the frequency range $\Delta f$.   At the same time from Eq.~(\ref{a3)}: $ \overline{x_{noise}^2} = \left|\chi (\omega)\right|^2 \frac{\overline{F_{noise}^2}}{m^2}$, and thus, one can obtain the spectral density of the noise displacement
\begin{equation}
\begin{split}
x_{th}(\omega) = \sqrt{\overline{x_{noise}^2}} = 
\left|\chi(\omega)\right|\alpha_{th}= \\
=\left(\frac{4k_b T\omega_n}{m\cdot Q \left(\left(\omega_n^2-\omega^2\right)^2+\left(\frac{\omega_n \omega}{Q}\right)^2\right)}\right)^\frac{1}{2}
\end{split}
\label{b12}
\end{equation}
which at the low frequency limit ($\omega \ll \omega_n$), where the $\chi(\omega)=\frac{1}{\omega_n^2}$, translates to  
\begin{equation}
x_{th}(\omega \ll \omega_n) = \frac{\alpha_{th}}{\omega_n^2}=\sqrt{\frac{4 k_b T}{m Q \omega_n^3}} =Const;
\label{b13}
\end{equation}
indicating flat response to the thermal noise within the low frequency limit.

\end{document}